\documentclass[]{interact}

\usepackage{rotating}
\usepackage{array}
\usepackage{graphicx}
\usepackage{subcaption}

\usepackage[natbibapa]{apacite}
\bibpunct[, ]{(}{)}{;}{a}{,}{,}

\usepackage{amsfonts}

\usepackage{hyperref}
\hypersetup{
    colorlinks = true,
    linkcolor=black,
    anchorcolor=black,
    citecolor=black,
    filecolor=black,
    menucolor=black,
    runcolor=black,
    urlcolor=black
}

\makeatletter
\def\maxwidth{\ifdim\Gin@nat@width>\linewidth\linewidth
\else\Gin@nat@width\fi}
\makeatother
\let\Oldincludegraphics\includegraphics
\renewcommand{\includegraphics}[1]{\Oldincludegraphics[width=\maxwidth]{#1}}

\theoremstyle{plain}

\theoremstyle{definition}

\theoremstyle{remark}

\usepackage[T1]{fontenc}
\usepackage[utf8]{inputenc}

\newcommand*\rot{\rotatebox{90}}

\usepackage{etoolbox}
\usepackage{tcolorbox}
\usepackage{array}

\tcbset{colback=gray!5!white,colframe=gray!85!black}

\AtBeginEnvironment{quote}{\vspace{3mm}\begin{tcolorbox}[leftrule=2mm,bottomrule=0mm,toprule=0mm,rightrule=0mm,boxsep=0mm,grow to right by=-3mm, grow to left by=-3mm]\small}
\AtEndEnvironment{quote}{ \end{tcolorbox}}

\author{
	\name{Wouter Groeneveld\textsuperscript{a}\thanks{CONTACT Wouter Groeneveld. Email: wouter.groeneveld@kuleuven.be}, Laurens Luyten\textsuperscript{b}, Joost Vennekens\textsuperscript{a}, and Kris Aerts\textsuperscript{a}}
	\affil{\textsuperscript{a}Department of Computer Science, KU Leuven, Leuven, Belgium; \textsuperscript{b}Department of Architecture, KU Leuven, Sint-Lucas Brussels and Ghent Campus, Belgium}
}
\begin{document}

\articletype{}

\title{Students' and Professionals' Perceived Creativity In Software
Engineering: A Comparative Study}

\maketitle

\begin{abstract}
Creativity is a critical skill that professional software engineers
leverage to tackle difficult problems. In higher education, multiple
efforts have been made to spark creative skills of engineering students.
However, creativity is a vague concept that is open to interpretation.
Furthermore, studies have shown that there is a gap in perception and
implementation of creativity between industry and academia. To better
understand the role of creativity in software engineering (SE), we
interviewed 33 professionals via four focus groups and 10 SE students.
Our results reveal 45 underlying topics related to creativity. When
comparing the perception of students versus professionals, we discovered
fundamental differences, grouped into five themes: \emph{the creative
environment}, \emph{application of techniques}, \emph{creative
collaboration}, \emph{nature vs nurture}, and \emph{the perceived value
of creativity}. As our aim is to use these findings to install and
further encourage creative problem solving in higher education, we have
included a list of implications for educational practice.
\end{abstract}


\hypertarget{introduction}{%
\section{Introduction}\label{introduction}}

\label{sec:intro}

This is the pre-print of the accepted publication for the Taylor \&
Francis European Journal of Engineering Education.

Software development is a field riddled with technical complexity and
constant change. Therefore, much research is devoted to these technical
aspects, while non-technical skills are given less attention: a recent
Delphi study has shown that expert software engineers in industry deem
creativity as a crucially important problem solving skill
\citep{groeneveld2020non}. Yet, creativity is still very much
underrepresented in software engineering (SE) education. For example, in
the ACM/IEEE Computer Science Curricula of 2020, the term
``communication'' is mentioned 84 times, while ``creativity'' and
``creative'' are collectively only mentioned 14 times \citep{cs2020}.
Furthermore, a curricula analysis by \citet{groeneveld2020soft} reported
that less than 5\% of the learning outcomes in computing-related courses
across European universities explicitly mention creativity. This shows a
skill gap between what industry experts think is important and what is
currently being taught in higher education.

Creativity research has a long history in the field of cognitive
psychology \citep{amabile1988model, torrance1972predictive}. The same
holds true for problem solving in the field of SE
\citep{robillard2005opportunistic, jalote2012integrated}. However, there
is still much to be done in order for the two fields to successfully
fuse. Creativity is typically a very vague concept to which everyone
gives their own interpretation and definition \citep{davis1999barriers}.
Furthermore, a recent systematic literature review concluded that:

\begin{quote}
\emph{``The research work on creativity in SE is scattered and scarce''
\citep{amin2017snapshot}.}
\end{quote}

The literature review of \citet{groeneveld2022creatively} summarizes how
creatively we are currently teaching and assessing creativity in SE
education. The authors conclude that most papers mention creativity only
in passing, often in one sentence. Sometimes, a section is devoted to
discussing its context, but often it is not used to support the
methodology of the paper or to interpret the results.

And yet, the core of SE is labeled as an outcome of human knowledge and
creativity \citep{bjornson2008knowledge}. Clearly, creativity is the
cornerstone of problem solving, especially in engineering, yet it is in
dire need of more attention.

\citet{kaufman2007creativity} identified three important aspects of
creativity. First, for an idea or implementation to be creative, it has
to be new or innovative. Second, it needs to be of high quality. Third,
the creative solution has to be relevant and appropriate to the task at
hand. However, it becomes increasingly difficult to come up with
completely novel ideas that nobody has thought of before. This is
especially difficult in mature fields, but also in the rapidly changing
field of SE. Someone who invents a new database model might be seen as a
visionary, while others who use the model in their daily programming
assignments are ``merely'' applying existing principles, even though
some developers might be applying the new model for the first time, for
them in a perhaps creative way. The difference between inventing it and
using it creatively is expressed as Big-C and Little-C
\citep{kaufman2009beyond}.

The three aforementioned aspects of creativity and the Little-C/Big-C
system are unfortunately never a perfect fit. More recently, psychology
researchers---including Kaufman---have been leaning towards a new
creativity theory: that of socio-culturality
\citep{glaveanu2020advancing}, where creativity is approached as a
contextual social verdict. Something is creative when someone else with
expertise and recognition says it is. This view of creativity highlights
the importance of context and perception of the construct. Therefore, we
wondered:

\begin{itemize}

\item
  \texttt{RQ1}: \emph{How do professionals perceive creativity in the
  field of SE?}
\item
  \texttt{RQ2}: \emph{How do students preceive creativity in the field
  of SE?}
\item
  \texttt{RQ3}: \emph{What are the similarities and differences in
  perception of creativity between both worlds?}
\end{itemize}

SE is a very broad discipline and requires systematic appliance of many
different techniques, ranging from programming, architectural and
interface design, maintenance of continuous deployment pipelines, to
juggling vague wishes of different stakeholders. To limit the scope of
this research, we will zoom in on the problem solving part of SE in
context of programming.

To answer question 1 and as a first step towards exploring the role of
creativity in SE and its relation to problem solving, we conducted four
focus group sessions, inviting 33 experts from four nationally and
internationally renowned companies, to creatively brainstorm about
creativity. This resulted in 399 minutes of transcripts, coded into 39
themes and grouped into seven domains: \emph{technical knowledge},
\emph{communication}, \emph{constraints}, \emph{critical thinking},
\emph{curiosity}, \emph{creative state of mind}, and \emph{creative
techniques}. However, one critical aspect remained unknown: how do
students perceive creativity?

Therefore, we conducted additional interviews, inviting 10 SE students:
5 undergraduates and 5 graduates. In this paper, we present the analysis
of both the professionals' and students' view of creativity in context
of problem solving during programming. In addition, and to answer
\texttt{RQ2} and \texttt{RQ3}, we thoroughly re-analyzed the transcripts
of the focus groups and conducted a comparative study. Our ultimate goal
with this research is threefold. First, to shed more light on the
socio-cultural context of creativity in SE. Second, to reveal important
differences in perception of creativity between students and
professionals. Third, to offer suggestions for the teaching community to
help close that gap.

The remainder of this paper is divided into the following sections.
Section \ref{sec:bg} outlines the background and related work, while
Section \ref{sec:method} describes the used methodology to approach the
interviews and comparative study. Next, in Section \ref{sec:results} and
\ref{sec:discussion}, we present and discuss the results. Section
\ref{sec:implications} summarizes implications for educational practice,
followed by possible limitations in Section \ref{sec:limits}, after
which Section \ref{sec:conclusion} concludes this work.

Part of this work, namely that related to \texttt{RQ1}, was previously
published at the 2021 ICSE-SEIS conference as
\citet{groeneveld2021exploring}.

\hypertarget{related-work}{%
\section{Related Work}\label{related-work}}

\label{sec:bg}

This section provides an overview of related work, both to better
identify creativity within engineering (education), and to explore
possible gaps between students' understanding of creativity and that of
professionals. We will highlight and summarize important published works
based on the aforementioned systematic literature reviews of
\citet{amin2017snapshot} and \citet{groeneveld2022creatively}.

\hypertarget{defining-creativity}{%
\subsection{Defining Creativity}\label{defining-creativity}}

Existing literature proves that pinpointing the boundaries of the
concept of creativity is very challenging: hundreds of definitions and
models of creativity exist across a plethora of domains, from
engineering to economics, design, linguistics, and social sciences. We
will highlight a few interpretations that stood out and can be easily
linked back to SE.

\citet{gero2000computational} approaches the creative design process by
comparing it to innovative and routine design. Routine design is defined
in computational terms as an activity which occurs when all necessary
knowledge is a priori available: a pre-established procedure can be
followed to come to a design solution. In contrast, he defines
non-routine design by two subgroups: innovative and creative design. In
innovative design, the values of the variables directing the procedure
to establish an outcome are placed outside their intended range. This
leads to design outcomes that are new but still belong to the same class
as their routine progenitors. In creative design one or more new
variables are introduced in the process leading to an all together new
class of design outcomes. Note that Gero's work stays conceptual:
identifying these variables or the perception of them is not mentioned.

According to the literature reviews of Section \ref{sec:intro}, popular
recurring creative models that stem from cognitive psychology are, among
others, Amabile's \emph{Expertise, Creative Thinking Skills, Motivation}
\citep{amabile1988model} and Mooney's 4P model (\emph{Process, Product,
People, Place}) \citep{mooney1963conceptual}. These models have been
studied before in context of SE, for instance by
\citet{amin2017snapshot}. Others combine multiple components the 4P
model, such as \emph{Place} and \emph{Process}, by the introduction of
project courses and project-based learning outside of conventional
university classrooms, to foster both technical and soft
skills---including creativity \citep{sedelmaier2014practicing}.

However, by limiting the exploration of creativity to borrowing and
translating an existing model, many of these studies try to push a
square peg in a round hole. Our intention is to first explore the
perception of creativity bottom-up and only then hark back to existing
models that might or might not show similarities. Furthermore, many of
these models are incompatible with the accepted idea of advancing
creativity theory and research to a socio-cultural concept, as reported
by \citet{glaveanu2020advancing}.

\hypertarget{perceiving-differences-in-creativity}{%
\subsection{Perceiving (Differences In)
Creativity}\label{perceiving-differences-in-creativity}}

\citet{catarino2019breaking} explored students' understanding of
mathematical creativity by sending out a survey to 128 engineering
undergraduate students and asking them ``What do you understand by
mathematical creativity?''. The result categories were \emph{involving
mathematics} (the application of mathematics in other areas; connections
between mathematical concepts), \emph{problem solving} (multiple or
original ways to solve a problem), and \emph{out of the box thinking}.
Perhaps this hints at certain domain-general aspects of creativity,
versus domain-specific aspects related to only mathematics or SE. The
authors conclude with a remark that, compared to other related works,
the perception of their surveyed first-year students might differ
substantially from that of graduate students.

\citet{bjorner2013academic} approached the perception of creativity
differently, by interviewing academic teachers instead of students. The
researchers were interested in how cross-disciplinary teachers perceive
and facilitate creativity. Again, the results point towards
problem-based learning and problem solving as an important element for
the creative process. The interviewed teachers had very different
definitions of creativity, but they all emphasized the social context
(in this case, the cultural learning environment) as an important
factor, both at macro and micro level. Teachers also frequently link
motivation to creativity. Bjørner and Kofoed end with a call for action:
teachers seem to lack proper support and tools to facilitate creativity
in education.

\citet{mohanani2017perceptions} investigated perceptions of creativity
in SE research and practitioners by comparing a systematic mapping of
research literature with 17 interviews of professionals. They discovered
some agreement---Kaufman's novelty and usefulness---but more importantly
identified differences in the way creativity is conceptualized,
measured, and improved. Their research sheds some light on the
research/practice gap but does not directly include the perception of
students.

To our best knowledge, no study asks both programming professionals and
students what they think creativity is like in context of problem
solving. Our contribution complements the above works, as will be
further discussed in Section \ref{sec:discussion}.

\hypertarget{methodology}{%
\section{Methodology}\label{methodology}}

\label{sec:method}

This section describes the approach used to answer \texttt{RQ1} to
\texttt{RQ3}. The study was executed in three phases: gathering data
from professionals using focus groups, gathering data from students
using semi-structured interviews, and comparing these findings.

\hypertarget{interviewing-professionals}{%
\subsection{Interviewing
professionals}\label{interviewing-professionals}}

\begin{figure}[h!]
  \centering
  \includegraphics{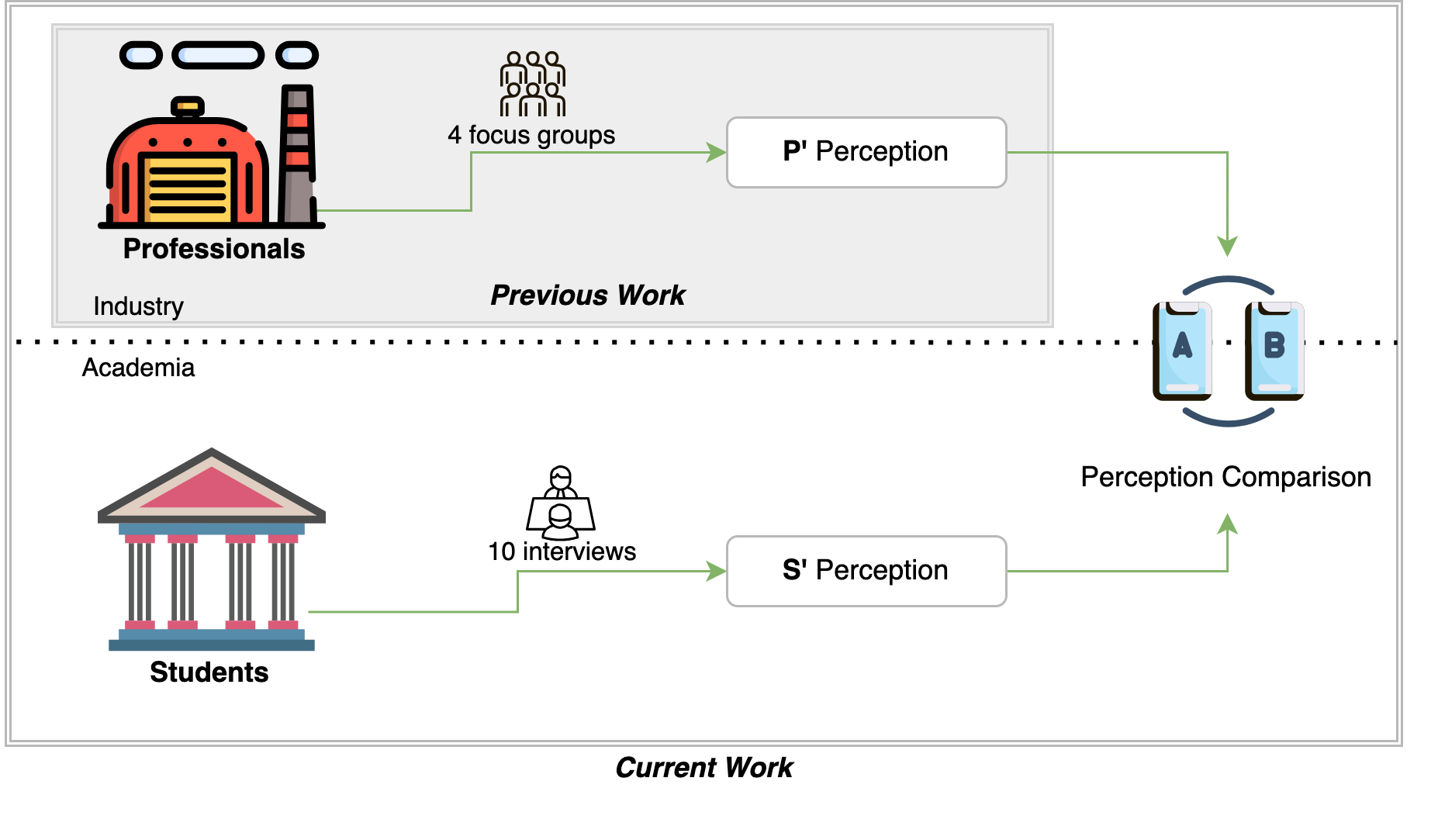}
  \caption{The context of previous and current works. \label{fig:context}}
\end{figure}

In 2021, we conducted four focus groups, inviting 33 SE experts from
four nationally and internationally renowned software development
companies. This study was published separately in
\citep{groeneveld2021exploring}. The transcripts were re-analyzed in
context of this paper. Figure \ref{fig:context} puts the previous and
this current work into context. We will briefly summarize the method
used for the focus group study below.

A focus group method was adapted for SE based on the work of
\citet{kontio2004using}. Focus groups, in which a small group of experts
brainstorm about a specific subject based on open questions of a
moderator, are well-suited to obtain feedback on new concepts and
generate ideas. To minimize participation friction, each session was
held at a separate company and consisted of employees of that particular
company. Studies have proven that this has the added advantage that
attendees feel comfortable since they know each other
\citep{fern2001advanced}. Participant selection was limited to these
having a technical role (e.g.~developers, programmers, software
architects, \ldots) with a minimum of six years of experience in the
field.

The conversation was facilitated by the first author, who has more than
11 years of experience as a SE himself. Familiarity with the topic at
hand could introduce bias, but \citet{kontio2004using} state that ``in
the field of SE, it is very important for the interviewer to have
extensive knowledge of the theme of the interview''. Sessions were
audio-recorded and transcribed, generating 399 minutes of transcripts,
on average 1.7h each. To minimize bias, they were cross-validated by the
second author, whose field of expertise lies outside of SE or computing.

To welcome opinions, perceptions, and ideas, open questions were posed.
In order not to influence the perception of the participants, we
refrained from providing a definition of creativity. We followed advice
of \citet{hove2005experiences} for conducting semi-structured interviews
in empirical SE research, which states that one should ``ask experience
questions, avoid `why' questions and questions with `yes'/`no'
answers''. The following questions were asked:

\begin{enumerate}
\def\labelenumi{\arabic{enumi}.}

\item
  \emph{What is the most creative thing you ever did related to
  programming?}
\item
  \emph{How can you see when a colleague is being creative, or not?}
\item
  \emph{What is the most creative project you ever worked on?}
\item
  \emph{How would you measure or assess creativity?}
\item
  \emph{Which creative techniques did you recently employ?}
\item
  \emph{What is the reason to be creative?}
\item
  \emph{Which environment do you need to be creative?}
\end{enumerate}

Additionally, to get the conversation on creativity jump-started,
participants were first asked to complete the sentences ``\emph{as a
developer, I am creative when I \ldots{}}'' and ``\emph{as a developer,
I am not creative when I \ldots{}}''. For each question, participants
were grouped in pairs to brainstorm for five minutes, after which ten
minutes of group discussion followed.

\hypertarget{interviewing-students}{%
\subsection{Interviewing students}\label{interviewing-students}}

The focus group study was a first step towards exploring the role of
creativity in SE. During academic year 2021--2022, we started a campaign
to enlist students with the intention to replicate the 2021 study using
exactly the same methodology but with different subjects: university
students. However, limited engagement of the students and planning
issues to get everyone together forced us to slightly revise the working
method.

Instead of organizing focus groups, we interviewed students one-on-one.
The questions were the same as the ones answered by the professionals
but slightly altered to better fit the context. For example, ``How can
you see when a \emph{colleague} is being creative'' became ``How can you
see when a \emph{fellow student} is being creative''. To keep the
conversation going, we asked follow-up and clarification questions about
responses we found intriguing (e.g., novel, vague, avoiding or in
contradiction to prior answers), again in accordance with Hove and Anda.
Although interviewees this time did not have the chance to brainstorm in
group before answering, we still gave them ample time to respond.

In total, 10 students were interviewed across 4 different universities
in the same North-European country, of which 5 undergraduate (first-year
students) and 5 graduate students (with four or five years in
university). To increase participation rates, students could secure a
video game voucher of 60 EUR with a 1/7 chance of winning. Neighboring
universities that provide a computer science course or an engineering
course with specialization in SE were asked to spread the announcement
in the form of a landing webpage via their channels. All interviews were
conducted in the students' native language to keep the conversation
fluid.

Two out of 10 interviewed students were female. All first-year students
were 18 year and all graduates either 22 or 23. Because
\citet{miller2014self} found no significant difference in
self-assessment creativity scores based on participant gender, ethnicity
or year in school, we decided not to consider student demographics as a
separate discriminant.

\hypertarget{analyzing-recorded-data}{%
\subsection{Analyzing recorded data}\label{analyzing-recorded-data}}

\label{sec:method:analyzing} Our intention was to treat, and thus
analyze, the data in the same way, for both groups of subjects
(professionals and students). The data was processed using the
qualitative coding techniques described by
\citet{onwuegbuzie2009qualitative} and \citet{richards2018practical}.
After audio recordings were transcribed, the transcript and written
notes were read multiple times to apply an open coding step. This was
done by the first two authors independently in order to let patterns
``bubble up'' in an emergent-systematic fashion as suggested by
Onwuegbuzie et al.~Next, notes were compared and cross-validated,
followed by an axial coding categorization step. After making multiple
revisions of mind maps and merging the results of all focus groups and
interviews, we finally arrived at 45 codes divided into 5 categories, as
visible in Table \ref{fulltable} in the appendix.

\begin{figure}
     \centering
     \begin{subfigure}[b]{0.48\textwidth}
         \centering
         \includegraphics{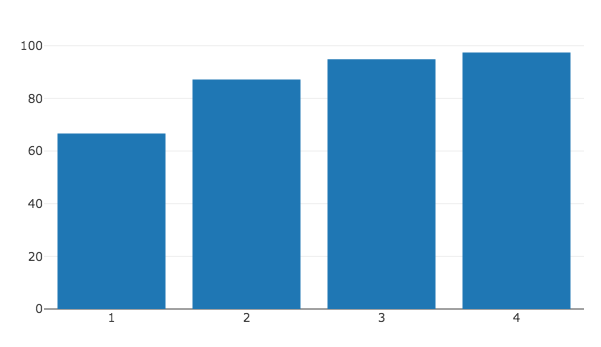}
         \caption{Focus groups with professionals.}
     \end{subfigure}
     \hfill
     \begin{subfigure}[b]{0.48\textwidth}
         \centering
         \includegraphics{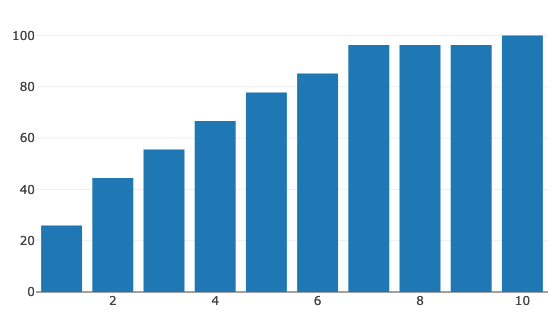}
         \caption{Interviews with students.}
     \end{subfigure}
        \caption{Indications of thematic data saturation: interviews (x-axis) projected on the percentage of identified codes (y-axis).}
        \label{fig:saturation}
\end{figure}

Thematic data saturation was calculated by counting the number of
discovered codes for each interview or focus group. As figure
\ref{fig:saturation} indicates, after the fourth focus group and the
tenth interview, less than 5\% new codes were introduced, which is a
good pre-set threshold according to \citet{guest2020simple}.

\hypertarget{comparing-findings}{%
\subsection{Comparing findings}\label{comparing-findings}}

Grounded theory dictates a bottom-up approach to identifying common
themes in qualitative data. While the application of this approach for
both datasets, as seen in Table \ref{fulltable}, proved to be a good
exercise, it ultimately failed to zoom in on the differences between
both groups. We felt that the axial coding step used to reduce the
number of subthemes is partially responsible for the loss of relevant
subtleties and context. For example, both professionals and students
mentioned working with (heterogeneous) groups, which can be seen as a
way to operationalize creativity---the ``how''---but also as a physical
setting---the ``where''. Furthermore, while it was mentioned by
students, the way in which they mentioned it made clear that they were
mostly theorizing (``I guess?''), compared to professionals who really
lean on collaborative efforts. Another identified subtheme, actively
seeking out input, appeared as a central theme in focus groups, but
merely as an aside in the student interviews.

To better highlight these subtle but very relevant systemic differences,
we reverted from a bottom-up approach to a more top-down approach,
zooming in on the differences that appeared in the first pass while
identifying codes bottom-up. While re-reading the manuscripts and
re-executing the coding steps a third time, we paid special attention to
the following concepts:

\begin{itemize}

\item
  The difference in environment in which the creativity is executed
  (professional vs a learning environment);
\item
  The awareness and application of creative techniques: by chance or on
  purpose;
\item
  Collaborative aspects of creativity and its perception;
\item
  The perceived role of nature vs.~nurture in the ability to be
  creative;
\item
  The perceived value of creativity in the profession or the project at
  hand.
\end{itemize}

In Section \ref{sec:results}, we will provide a brief overview of the
bottom-up findings, before zooming into the comparative similarities or
differences, described in Section \ref{sec:discussion}.

\hypertarget{bottom-up-results}{%
\section{Bottom-up Results}\label{bottom-up-results}}

\label{sec:results}

Tables \ref{tab:metadata-p} and \ref{tab:metadata-s} contain summaries
of gathered focus group and interview data. In total, 43 participants
generated 89.462 words in 652 minutes worth of transcripts. On average,
a focus group session with professionals---of on average 8
participants---lasted 100 minutes, while an interview session with
students took 25 minutes.

The full data set of anonymized transcripts and source code used to
process these are available at \url{https://doi.org/10.48804/3FHVRC}.

\begin{table}[h!]
  \centering
  \small
  \caption{Focus group metadata of 33 professionals, totaling 52.329 words and 399 minutes of data. Theme distribution rates are also visible in Figure \ref{fig:saturation}. \label{tab:metadata-p}}
  \begin{tabular}{l | r r r r}
    Description & P1 & P2 & P3 & P4 \tabularnewline
    \hline
    \hline
    Duration & 105m & 99m & 100m & 95m \tabularnewline
    Participants & 7 & 14 & 7 & 5  \tabularnewline
    Words & 13 454 & 12 296 & 13 492 & 13 087  \tabularnewline
    Theme distr. & 67\% & 87\% & 95\% & 100\%  \tabularnewline
  \hline\end{tabular}
\end{table}

\begin{table}[h!]
  \centering
  \small
  \caption{One-on-one interview metadata of 10 students, totaling 37.133 words and 253 minutes of data. Theme distribution rates are also visible in Figure \ref{fig:saturation}. \label{tab:metadata-s}}
  \begin{tabular}{l | r r r r r r r r r r}
    Description & S1 & S2 & S3 & S4 & S5 & S6 & S7 & S8 & S9 & S10 \tabularnewline
    \hline
    \hline
    Duration & 23m & 25m & 22m & 22m & 24m & 31m & 24m & 24m & 23m & 35m \tabularnewline
    Words & 3 321 & 4 211 & 2 722 & 3 028 & 3 533 & 4 802 & 3 812 & 2 713 & 4 247 & 4 690 \tabularnewline
    Theme distr. & 26\% & 44\% & 56\% & 67\% & 78\% & 85\% & 96\% & 96\% & 96\% & 100\% \tabularnewline
  \hline\end{tabular}
\end{table}

The following subsections provide a summary of the emerged subthemes and
categories while bottom-up analyzing the transcripts of professionals
and students, as also presented in Table \ref{fulltable}. Themes are
marked in bold. We will zoom in on the top-down approach in Section
\ref{sec:discussion}.

\hypertarget{when-are-you-not-creative}{%
\subsection{When are you (not)
creative?}\label{when-are-you-not-creative}}

Perhaps unsurprisingly, most professionals and students agree that
creativity is needed to come up with a \textbf{unique solution} to a new
problem. This problem should be \textbf{challenging}, yet at the same
time not \textbf{too difficult}. If it is too easy, it is considered as
mundane and boring, and creativity rapidly wanes. If the problem is
perceived as too difficult---for instance if the API or programming
language is completely new---participants first try to learn and follow
existing rules before attempting to tackle it creatively. According to
\citet{csikszentmihalyi1997flow}, just the right amount of difficulty is
important to achieving a creative flow-like state.

Students put heavy emphasis on \textbf{freedom of choice}: they report
being the most creative if they have full control over the assignment
such as choice of framework and implementation method. On the other
hand, when a course project imposes a certain technology and is less
open, they consider it to provoke less creativity. This goes against the
grain of professionals' coding environments that are usually filled to
the brim with imposed constraints. We will explore this more in detail
in a later section.

For some participants, working on \textbf{something visual} like the UI
design or making drawings on a whiteboard is perceived as being more
creative. For others, their creativity sparks when \textbf{performance}
is a factor, again hinting at the importance of both imposed and
self-imposed constraints in creative problem solving.

One student jokingly mentioned that \textbf{sleeping well} and being in
the right mood also determines when they are more or less creative. This
lines up with the literature that suggests that sleep does indeed
facilitate creativity \citep{ritter2012good}.

During the warm-up questions, some professionals attempted to define
creativity:

\begin{quote}
\emph{When someone can come up with an elegant solution to a previously
unsolved non-trivial problem.}
\end{quote}

This turned out to be quite challenging, as the definition was almost
immediately heavily disputed. All students and professionals did agree
on the fact that creativity is inherently personal. Some students
suggested it was subjective and depending on the person evaluating the
creativity, which lines up neatly with the attention that is currently
paid in cognitive psychology to the socio-cultural context of
creativity.

\hypertarget{why-should-you-be-creative}{%
\subsection{Why should you be
creative?}\label{why-should-you-be-creative}}

When asked why creativity matters, interviewees unanimously answered
because \textbf{quality matters} and can only be achieved by staying
longer with the problem, coming up with alternatives, poking at the
system from different viewpoints, again \textbf{overcoming constraints}
that obstruct the path towards creation.

Besides conventional answers, such as ``because \textbf{innovation
requires creativity}'', both students and professionals hint at the
\textbf{personal satisfaction} of approaching coding problems
creatively. Without creativity, things become boring fast, according to
respondents:

\begin{quote}
\emph{In a sense, creativity can work therapeutically. I wouldn't be
here if this work wasn't creative.}
\end{quote}

Recent work by \citet{graziotin2014happy} revealed that happy
programmers perform better, and thus might be more creative. Some
participants went as far as saying that creativity \textbf{is part of
their personality}, of who they are: a \emph{Homo Faber}, Umberto Eco's
concept of the urge to use tools and manipulate technology to create and
shape the world around us \citep{eco1989open}.

Assembly line work---or in programmers terms, generating getters and
setters and doing the plumbing setup work of a project---was frequently
mentioned as being horribly dull. Still, some professionals like to pick
up mundane tasks to trigger later creative ``aha'' moments, while
students rejected this kind of work all-together.

The ``why'' theme again hints at dissimilarities between students' and
professionals' opinions: professionals are tuned to more iterative
software development processes that facilitate \textbf{fast feedback}
and \textbf{client-oriented thinking}. Within the safe walls of the
university, this is completely absent and can only be partially
emulated. Professionals also more openly \textbf{admire others' creative
work} which in turn sparks their creative skills. One student did
mention that seeing other students at work on a similar exercise can
help sparking creativity. However, for professionals, finding a creative
muse goes beyond similar work: some even marveled at the engineering
work of suspension bridges or highway interchanges.

\hypertarget{what-is-creativity-like}{%
\subsection{What is creativity like?}\label{what-is-creativity-like}}

Besides the \textbf{subjective} nature of creativity that was brought up
multiple times, when we asked interviewees how creativity could be
recognized and measured, many thought it could be observed by
\textbf{following the reasoning} of others after asking them an
open-ended question. If they deliberately think of alternatives,
critically evaluate options, sometimes backtrack, sometimes zoom in or
zoom out, then their (software development) thought process is deemed as
creative.

Both parties also identified the conventional answer: assessing
\textbf{divergent thinking} by counting the number of solutions that get
brought up. The more solutions, the more creative. In essence, this is
Torrance's Test of Creative Thinking \citep{torrance1972predictive},
that contemporary creativity research is moving away from
\citep{glaveanu2020advancing}, as divergent thinking is only one aspect
of creative problem solving.

Instead of trying to judge the creativity level of the development
process, interviewees suggested to \textbf{evaluate the end result}: the
product itself. The evaluation of a product as judged by an expert group
is called Amabile's Consensual Assessment Technique
\citep{amabile1982social}, and according to the literature reviews of
Section \ref{sec:intro}, it is an often used technique in SE education.

Both professionals and students mentioned \textbf{elegance in code}, but
both interpret this differently: students focus much more on complexity,
perhaps to show off their skills as part of the assignment, while
professionals put emphasis on creative clean code (and the many unit and
integration tests that come with it) to make long-term maintenance much
more bearable.

Only professionals mentioned \textbf{sharing responsibility} as an
integral part of creative problem solving in SE. Of course, the culture
and work environment differs radically compared to SE students in higher
education. A detailed comparison of different social and cultural
creative contexts is described in Section \ref{sec:discussion}.

\hypertarget{how-can-creativity-be-induced}{%
\subsection{How can creativity be
induced?}\label{how-can-creativity-be-induced}}

The most striking difference between both study groups is the number of
creative techniques that are brought up during interviews and focus
groups. Experienced programmers clearly are at an advantage here,
mentioning techniques such as \textbf{shifting angles} (going back and
forth between zoomed-out software architecture and zoomed-in code at
method level) and \textbf{rubber ducking} (literally arguing with a
rubber duck---yourself---when stuck to force a shift in perspective).
Some techniques such as quickly drafting a potential solution in
\textbf{pseudocode} and \textbf{dividing} a problem into small
sub-problems were mentioned by students but not by professionals. We
argue that they are so commonplace in industry that they are hardly
worth mentioning at all, while for students, they are still a fairly
novel practice.

Both parties mentioned getting \textbf{out of your comfort zone} and
\textbf{daring to ask questions}: if you are not open to learning new
things, your creativity will be barely fueled to keep on going. Feedback
and collecting information is important, both from people (mentioned
more often by professionals) and in the form of read-only information
found in books, websites, or other sources (mentioned more often by
students). One professional summarized it neatly as follows:

\begin{quote}
Creativity is the brew of different inputs.
\end{quote}

The concept of \textbf{continuous learning} is well-known within the
world of SE \citep{rodrigues2019continuous}, yet still fairly new to
students. They are still steeped in a very supportive learning
environment, while in industry, it takes deliberate practice to keep on
being involved in learning, making it a more important factor for
creative success.

\hypertarget{where-are-you-the-most-creative}{%
\subsection{Where are you the most
creative?}\label{where-are-you-the-most-creative}}

All interviewees emphasize the need for a \textbf{supportive
environment} in which mistakes are not immediately punished. For
students, this means dealing with open assignments that do not have a
one size fits all answer, while for professionals, this means management
that allows room for creative experiments.

Some programmers seemed to prefer working with \textbf{music} to have
easy access to Csikszentmihalyi's \textbf{flow state of mind} while
others prefer a \textbf{quiet} environment. SE students interpret that
quiet environment as literally \textbf{being alone} in front of a PC and
although both study groups mentioned \textbf{working in group}, this was
much more ingrained in the professionals' answers.

Generating creative ideas does not need to happen exclusively clutched
in front of a screen. \textbf{Taking a walk} or engaging in other
activities like sports to let the mind wander and subconsciously think
about a complex problem was mentioned by many respondents of both
groups. Existing research such as that of \citet{colzato2013impact}
supports the notion that physical movement enhances divergent thinking.
The ``aha moment'' usually does not occur when writing code but while
doing something else. In fact, most of the thinking work can be done
somewhere else, as one professional stated:

\begin{quote}
\emph{Most of my creative work happens in the car. When I'm at work, all
I have to do is type out the solution in my head.}
\end{quote}

Interestingly, professionals seemed to deliberately trigger these
moments while students recognize them when they occur but fail to
understand and apply it as a creative technique.

In the next session, we will discuss these differences in detail with
the help of the results of our top-down approach.

\hypertarget{top-down-results}{%
\section{Top-down Results}\label{top-down-results}}

\label{sec:discussion}

The previous section presented the results of a thorough bottom-up
analysis of the transcripts that revealed the when, why, what, how, and
where of creativity as an important factor in problem solving for
software engineers. As indicated by the interview occurrence columns in
the appendix, professionals and students did not always agree. This
section zooms in on those details and highlights both subtle differences
in interpretation and bigger gaps in the perception of creativity. We
end the discussion with a set of implications for educational practice.

\begin{figure}[h!]
  \centering
  \includegraphics{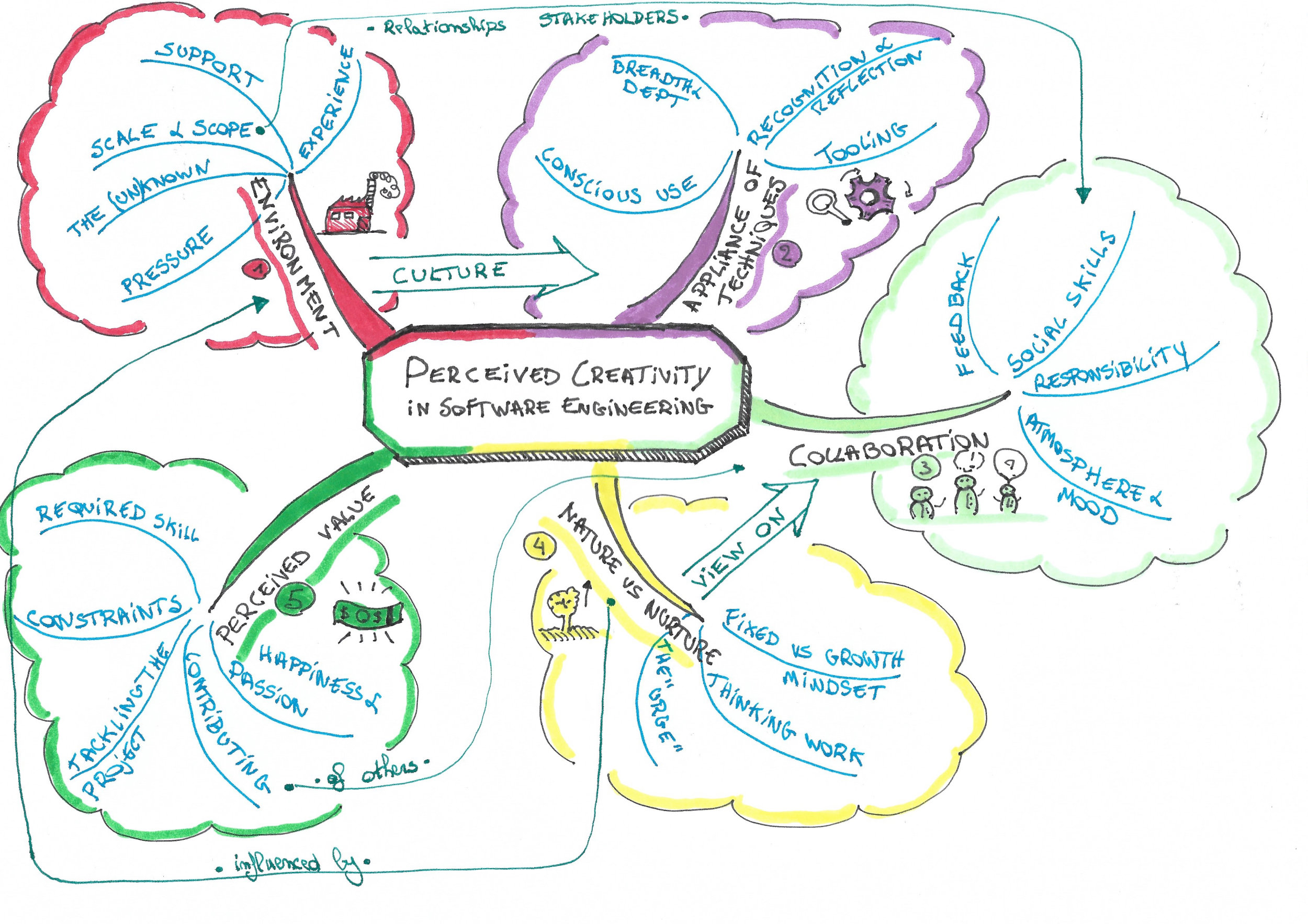}
  \caption{A mind map of the different perceptions of creativity in SE between both study groups. \label{fig:mindmap}}
\end{figure}

The mind map in Figure \ref{fig:mindmap} summarizes the five central
themes used in this top-down approach. It is important to note that
these themes are highly interconnected, as emphasized by the arrows in
the mind map. For instance, the way you interpret creativity alters your
view on collaboration, or the environment that determines the
company/university culture influences the application of certain
creative techniques. We agree with Nerubasska and Maksymchuk: creativity
is \emph{systemic} \citep{nerubasska2020demarkation}.

\hypertarget{environmental-differences}{%
\subsection{Environmental differences}\label{environmental-differences}}

Professionals and students have to exert their creativity in very
different environments, of which especially the \textbf{scale and scope}
of both the projects and everything that surrounds these is vastly
different. For instance, the much larger development environment in
industry includes:

\begin{itemize}

\item
  More context-switching between debugging, restructuring, and writing
  new code;
\item
  DevOps activities such as continuous integration and distributed
  deployment;
\item
  More emphasis on (automated) unit and end-to-end testing;
\item
  Long-term legacy software support and possibly data migration steps to
  undertake;
\item
  Much more frameworks and APIs to juggle;
\item
  More data, logs, and requests to process;
\item
  The involvement of many stakeholders that requires more interaction;
\item
  \ldots{}
\end{itemize}

To a professional software engineer, creativity isn't just limited to
writing code or fixing problems in code: it happens both at micro-level
in code and at macro-level in the software architecture. As typical
university assignments are very limited, even if they are open ended to
help spark creativity \citep{vandegrift2015supporting}, this part of
creative problem solving is virtually unknown to SE students.
Furthermore, the speed at which some frameworks come and go also
requires a certain critical and creative judgement during mid to long
term software development planning. Some professionals even mentioned
dealing with patents. THe experts testified that the more
\textbf{experienced} you are working in these environments, the easier
it becomes to express yourself creatively.

Another environmental difference is working with the \textbf{(un)known}.
Professionals state that, many times, clients do not really know what
they want, requiring a critical mindset, asking many questions, and
short feedback cycles. We found that many students did not dare to
question the question, which is a component to creativity
\citep{groeneveld2021exploring}. In a sense, problem solving in SE is
related to ``wicked'' problems in design thinking, which Buchanan
defines as ``\emph{as a class of social system problems which are
ill-formulated, where the information is confusing, where there are many
clients and decision-makers with conflicting values, and where the
ramifications in the whole system are thoroughly confusing}''
\citep{buchanan1992wicked}. Wicked problems require creativity to
successfully tackle.

Typical student projects exhibit some form of \textbf{pressure}, but
this is usually limited to time pressure. Enterprise software
development projects come with many levels of pressure: budgeting, time,
expectations from peers and management, having to support existing
legacy systems, \ldots{} Some of these can be classified as constraints
which we will explore in one of the following subsections.

Creativity can only happen if the environment \textbf{supports it}: are
peers and managers supportive? Does the company culture allow for a bit
of play that might spark new ideas? Students interpreted support as
coming from their friends and family, while professionals mention the
bigger picture. For students, most projects have only one purpose:
passing the course. Some admit to trying to optimize their grades. The
fragmentation of typical SE curricula and the lack of feedback during
and after the project does not work in favor of creativity.

\hypertarget{the-awareness-and-application-of-creative-techniques}{%
\subsection{The awareness and application of creative
techniques}\label{the-awareness-and-application-of-creative-techniques}}

Creative techniques have to be \textbf{used consciously}. In the
professional focus groups, one of the central themes was often
consciously choosing when \emph{not} to be creative: when writing a
script that only gets executed once or when the solution can be found by
repeating a previous practice and not time and time again questioning
the architecture or trying to reinvent the wheel. None of the
interviewed students mentioned this. A professional summarized this
cleverly:

\begin{quote}
\emph{You {[}only{]} have to be creative when it is time to do so.}
\end{quote}

Related to the conscious use of creativity is the \textbf{recognition of
and reflection on} these creative actions. The context of the student
interviews revealed that they might be capable of recognizing the
moments, but not yet of actively capitalizing on them. One student said:

\begin{quote}
\emph{I'll admit, I've never really looked for creativity. As is often
the case, either {[}the solution{]} works, or it doesn't. Most of the
time, I just sit here at my desk.}
\end{quote}

Also, because of the much larger environmental scope, the creative
techniques that professionals employ clearly showcase more
\textbf{breadth and depth}. Students seem to be very much focused on the
technical complexity of algorithms and the code on a micro level, while
professionals mention a much more diverse set of creative techniques
they utilize daily: switching between refactoring and redesigning,
zooming in and out, visualizing problems, adding in a so-called
``deliberate hack'', taking edge cases into account when writing tests,
\ldots{}

One of those examples that stood out was \textbf{tooling}. For SE
students, tooling meant again getting to work with technical complexity:
being creative by designing their own tooling systems. Interestingly,
professionals use tools (e.g.~editors and IDEs) to get rid of the
unnecessary clutter and be able to focus on the essence: the coding
problem at hand.

The lack of breadth and depth causes students to stop pushing forward by
switching to other creative techniques when one fails to work. One
student did mention writing many unit test cases to split a problem in
small sub-problems, but did not mention the other potential creative
advantage of testing that was mentioned by almost every professional:
poking at the problem from multiple angles to arrive at an airtight
implementation. This again signifies a lack of practical experience,
which is of course to be expected.

\hypertarget{perceived-collaborative-aspects-of-creativity}{%
\subsection{Perceived collaborative aspects of
creativity}\label{perceived-collaborative-aspects-of-creativity}}

As underscored by professionals:

\begin{quote}
\emph{For creativity, communication is key.}
\end{quote}

To them, various \textbf{social skills} to help deal with the
communicative and collaborative aspects of creativity are of importance
as well, especially in a professional environment where different
stakeholders have different expectations of both developers and the
project in development. Knowing how to approach others is key to getting
the most creative ideas out of that interaction.

``\emph{The ability to incorporate \textbf{feedback} of others in your
solution-oriented thinking is of course an important part of the
creative process}'', testified a professional. The importance of
feedback is echoed by students, however, their feedback is much more
limited. For some assignments, depending on the course or the teaching
staff, they do not receive any feedback at all, severely hampering
reflection on the creative process. Students also seemed to prefer
getting inspired by other work through books or online resources instead
of directly via discussions with their peers, hinting at a more
individual development process. Students do mention that working in a
diverse group will likely trigger more creativity, but the way they
bring it up seems to indicate this as being purely theoretical
knowledge:

\begin{quote}
\emph{People with a different background that might have followed
different courses that I didn't take might possibly think about things I
didn't think of. So I guess that's where the creativity probably
surfaces then. I think.}
\end{quote}

An existing body of literature indicates that interdisciplinary SE
projects are nothing new, although they almost never focus on improving
the creativity by taking advantage of the heterogeneous student groups.
In one study, interdisciplinary computing classes are said to improve
``soft skills'' of the students, but creativity was not mentioned
\citep{carter2014interdiscp}.

Professionals also think it is important to be given \textbf{shared
responsibility} to execute the planned work. This creates a unique
\textbf{atmosphere} that, according to our participants, amplifies
creativity. This sense of responsibility is absent with students: their
only responsibility is submitting the work on time to hopefully pass and
eventually graduate. Especially when closely working together,
atmosphere and collegiality does have an impact on mood, which has been
proven before to greatly affect the willingness to be creative
\citep{graziotin2014happy}.

Creativity can be contagious. Professionals often strategically place
newcomers next to creative enthusiasts to help them quickly get up to
speed. Students acknowledge this, but again, it sounds like theory that
was never put into practice:

\begin{quote}
\emph{{[}\ldots{]} I didn't think of that yet, usually that should also
lead to new insights for yourself, I guess, when you see someone else
doing something creative.}
\end{quote}

\hypertarget{nature-vs-nurture-the-creative-mindset}{%
\subsection{Nature vs nurture: the creative
mindset}\label{nature-vs-nurture-the-creative-mindset}}

While re-reading the transcripts, one thing that stood out was the
\textbf{Fixed vs.~Growth mindset} of students vs professionals.
Cognitive psychologist Carol Dweck describes a Fixed mindset as one that
is rigid: either people are capable of something, or they are not; while
a Growth mindset acknowledges the possibility to nurture skills
\citep{dweck2008mindset}.

When it comes to creativity in programming, we observed primarily a
Fixed mindset when interviewing students and primarily a Growth mindset
when interviewing professionals. For professional software developers,
continuous learning is part of their job, as every single component in
their environment rapidly changes. They regard creative problem solving
as a skill that can be learned, just like debugging or database
maintenance. On the other hand, students tend to categorize people into
ones that are very creative and ones that are not:

\begin{quote}
\emph{There are people who are spontaneously creative. and who don't
think about it, while I do, I do introspect when I want to make
something, and think about my work, so I'm consciously working on that.}
\end{quote}

That person continued to describe behavior of friends who he thought
were more creative, and fellow students who in his mind weren't. While
``consciously working on that'' sounds like a Growth mindset, in his
mind, some people are capable of it, while others are not. It did not
occur to him that this can be trained. A few students mentioned
creativity was related to intelligence and seemed to suggest that
intelligence is also very much a fixed affair. Students' rigid mindset
attitude towards creativity also surfaced in other focus group studies
\citep{katz2017engineering}.

A ``sudden'' creative idea is the result of a lot of \textbf{thinking
work}. ``\emph{Being creative involves a lot of hard work}'', testify
several professionals. They continue by saying that hard work is best
alternated with deliberate breaks or periods of mundane work to let the
unconscious do its work, ultimately triggering that ``aha'' moment.
Ideas never occur in a sudden daze of brilliance. As for students, they
recognize that moment, for example during a long walk or another sports
activity, but do not deliberately employ these techniques to let ideas
bubble up more often. Some students even admitted to not ``catching''
the idea---for instance by writing it down---and promptly forgetting it.
All students found repetition to be very much not creative, while
professionals sometimes insert periods of repetitive work to explicitly
trigger something else.

Many professionals described their creative endeavors as the result of
their \textbf{creative urge}: the need to do something creative, Eco's
\emph{Homo Faber} \citep{eco1989open}. This was less distinctly present
with students, although they also all acknowledge the need to be
creative, but interpret that mostly as having creative freedom.

\hypertarget{the-perceived-value-of-creativity}{%
\subsection{The perceived value of
creativity}\label{the-perceived-value-of-creativity}}

Both groups agree that creativity is a \textbf{required skill}: it is
needed to find and solve non-trivial programming problems, to keep on
digging while bug fixing, \ldots{} We summarized this as
\textbf{tackling the project}.

An important aspect of problem solving is dealing with
\textbf{constraints}. Compared to students, professionals face a
plethora of constraints: time, budget, code, legacy, and so forth. For
many professionals, self-imposed constraints are just as important as
imposed constraints, as they usually trigger creative thinking.
Self-imposed constraints are an often-used technique in art and
interaction design \citep{biskjaer2011self}. For many professionals,
working on a \emph{brownfield} project---where existing software
architectures have to be taken into account---was considered more
creative than a \emph{greenfield project}---where a team can start from
scratch and use the latest tech---although this was sometimes debated.

Students do mention constraints but in a much more limited sense. For
most students, constraints work against creativity. One student noted
that they preferred dynamic programming languages over static ones
``\emph{where everything is set in stone}'', failing to see that exactly
those limited environments breed the most creative solutions. Every
student preferred a greenfield project, working from scratch, to be in
full creative control. Only one student---who turned out to be working
part-time for two years in the SE industry---mentioned self-imposed
constraints.

Expressing oneself creatively can improve well-being, as expressed by
several professionals and described in Section \ref{sec:results}. For
many professionals, writing code is seen as a \textbf{passion} or a
craftmanship: pride is taken in delivering something of high quality,
both on the outside (for the end users) as in the inside (clean code for
the software maintainers). Students also express their fondness of
creativity---that is, when an assignment allows some degree of creative
freedom. Of course, for them, there is less incentive to write clean
code beyond getting a grade: most code written in university is
throwaway code.

Some students enthusiastically talked about their hobby coding projects
where they are in complete control and can \textbf{contribute} to
society by solving a problem that others also encounter. That ``complete
control'' again highlights students' preference for working alone to
cling to that precious freedom. Some students even expressed their
aversion for working together as compromises have to be made which might
muddle their creative vision.

\hypertarget{implications-for-educational-practice}{%
\section{Implications for educational
practice}\label{implications-for-educational-practice}}

\label{sec:implications} We are convinced that some of the identified
differences in the perception of creativity can lead to actionable
advice such as explicitly introducing creative problem solving
techniques in SE courses. To summarize the results and discussion of the
previous sections, we present a set of implications for educational
practice that can be used to introduce and improve creativity in the
classroom.

\begin{enumerate}
\def\labelenumi{\arabic{enumi}.}

\item
  Even though students perceive constraints to be a helpful part of
  constraints, as visible in Figure \ref{fig:mindmap}, they do not know
  how to bend them to your advantage. \newline \emph{Help students
  discover the benefits of (self-imposed) constraints, for example by
  introducing brownfield projects.}
\item
  Students are used to working with small projects that do not
  faithfully resemble a typical SE project. \newline \emph{Keep the
  scale and scope of a typical SE project in mind when designing
  assignments, for example by involving more (imaginative)
  stakeholders.}
\item
  Students sometimes focus too much on getting a grade, thereby
  explicitly avoiding a creative approach. \newline \emph{Encourage
  creative explorations in problem solving, even if they do not lead to
  practical solutions, for example by focusing less on grading the end
  result and more on the creative process.}
\item
  Feedback happens too little and can come across as too official.
  \newline \emph{Plan for frequent moments of feedback, before, during,
  and after the submission of the final assignment.}
\item
  Students often don't know when to be creative and when not to be.
  \newline \emph{Help students use a diverse set of creative techniques,
  including when \emph{not} to be creative, for example by presenting
  multiple unique SE problem cases.}
\item
  Many students think that either they're creative, or they're not.
  \newline \emph{Show students that creativity is not a static personal
  characteristic but a skill that can evolve.}
\item
  Classic course assignments push students towards individualism.
  \newline \emph{Emphasize the power of collaborative idea generation
  and iterative creativity, for example by introducing and actively
  coaching more group work.}
\end{enumerate}

\hypertarget{limitations}{%
\section{Limitations}\label{limitations}}

\label{sec:limits} This comparative study sought to answer the research
questions using qualitative methods. As we have discussed, creativity is
a subjective matter, and opinions of experts and students vastly differ,
making it challenging to draw general conclusions. Repeating our study
with other participants, for example from different companies or
universities abroad, might produce different results. However, by
carefully cross-validating data and adhering to the best-practices of
qualitative research, as explained in the methodology section, we
believe that we have mitigated the biggest risks. The data saturation
rate indicates that the 4 focus groups and 10 interviews provided enough
unique data to proceed with encoding and analyzing the results.

The interviews with students did not follow the exact same procedure as
the focus groups. One-on-one interviewees might experience more pressure
in trying to quickly answer questions compared to brainstorming in
group, resulting in possibly short-sighted answers. To minimize this
effect, we compared encountered codes in student interviews with those
from the focus groups and grouped them accordingly.

Another possible limitation to this study is selection bias.
Professionals mentioned this as well: only the people that were
intrinsically motivated to talk about creativity joined our focus
groups. We intentionally selected on this intrinsic interest to increase
the relevance of answers for answering \texttt{RQ1} and \texttt{RQ2}.
Still, it remains uncertain how much creativity is applied on a
day-to-day basis in the field of SE, or how many students deliberately
make use of creative techniques when working through their assignments.

Lastly, one of our initial aims for this research was to discover
potential differences in the perception of creativity between
undergraduate and graduate students. While it was clear---and perhaps
expected---that students with some work experience started to lean
towards the opinion of professionals, there was not enough data to make
any claims regarding an evolving view of creativity as students mature
during their education. Future research might shed more light on this.

\hypertarget{conclusion}{%
\section{Conclusion}\label{conclusion}}

\label{sec:conclusion} This paper sought to explore the perception of
creativity in the field of SE. For this, we gathered transcripts from
two study groups: 33 experts in the industry grouped into 4 focus group
sessions and 10 students in higher education who major in SE. Five
themes emerged from a bottom-up qualitative code analysis: \emph{when to
be creative} (9 subthemes), \emph{why should you be creative} (9),
\emph{what is creativity like} (7), \emph{how to be creative} (11), and
\emph{where does it manifest} (9).

Next to these themes, a follow-up top-down analysis zoomed in on the
similarities and differences between both study groups. The comparison,
yielding 21 new subthemes, was summarized using the following five
themes: \emph{the creative environment} (5 subthemes), \emph{application
of techniques} (4), \emph{creative collaboration} (4), \emph{nature
vs.~nurture} (3), and \emph{the perceived value of creativity} (4).

This paper contributes to the engineering education research community
by helping identify the gap between SE industry and academia. In future
work, we will experiment with various creative interventions and measure
the effectiveness of creativity teaching techniques. However, first and
foremost, we must focus on helping evolve students' Fixed mindset of
creativity into a Growth mindset.

\hypertarget{acknowledgements}{%
\section{Acknowledgements}\label{acknowledgements}}

We would like to thank the participants of both study groups for their
creative discussions, insights, and suggestions on creativity in
software engineering. Thank you all very much!

\hypertarget{appendix}{%
\section{Appendix}\label{appendix}}

Table \ref{fulltable} contains the identified codes that were grouped
during the first analysis pass as described in Section
\ref{sec:method:analyzing}, including a presence marker of the code in
either the transcripts of the professionals and/or the students.

\begin{table*}[h!]
\footnotesize
\begin{center}
\begin{tabular*}{\textwidth}{l p{0.65\textwidth}|>{\centering}p{0.10\textwidth}|>{\centering\arraybackslash}p{0.10\textwidth}}
\\
\toprule
 & \textbf{Subtheme}
 & \textbf{P. Occ.}
 & \textbf{S. Occ.}
\\
\midrule
  & Not yet a solution, a unique/new problem    & \checkmark & \checkmark \\  
  & Slept well, in the mood    &  & \checkmark \\  
  & A challenge that dictates thinking    & \checkmark & \checkmark \\
  & Not too difficult    &  & \checkmark \\  
  & Linking with other domains    & \checkmark & \checkmark \\  
  & Creating something visual    & \checkmark & \checkmark \\    
  & Freedom of choice    & & \checkmark \\
  & Contribute something to society    & & \checkmark \\
\rot{\rlap{\textbf{When? (9)}}} & Performance matters    & \checkmark & \checkmark \\
\midrule
  & Quality matters    & \checkmark & \checkmark \\  
  & Overcoming constraints     & \checkmark & \checkmark \\  
  & Client-oriented thinking    & \checkmark &  \\  
  & Innovation/renovation is needed    & \checkmark & \checkmark \\  
  & Because otherwise things get boring    & \checkmark & \checkmark \\    
  & Personal satisfaction    & \checkmark & \checkmark \\  
  & That's just the way I am    & \checkmark &  \checkmark \\  
  & Fast feedback    & \checkmark &  \\    
\rot{\rlap{\textbf{Why? (9)}}} & Admiration for other creative work    & \checkmark &  \\  
\midrule
  & Subjective; taste-dependent    & \checkmark &  \checkmark \\    
  & Studying their reasoning    & \checkmark & \checkmark \\    
  & Counting the amount of different solutions    & \checkmark & \checkmark \\    
  & Judge the end result    & \checkmark & \checkmark \\    
  & Check for elegance in code    & \checkmark & \checkmark \\    
  & Shared responsibility    & \checkmark &  \\    
\rot{\rlap{\textbf{What? (7)}}} & Body language; surprise    & \checkmark & \checkmark  \\    
\midrule
  & Pseudocode rendering    &  & \checkmark \\    
  & Divide and conquer    &  & \checkmark  \\      
  & Drawing diagrams, domain modeling    & \checkmark & \checkmark \\    
  & Continuous learning    & \checkmark &  \\    
  & Actively seeking out new input/inspiration    & \checkmark &  \checkmark \\    
  & Discussions/brainstorming (on whiteboard)    & \checkmark & \checkmark  \\    
  & Exploring different angles/bird's eye view    & \checkmark &  \\    
  & Ask yourself questions: rubber ducking    & \checkmark &  \\    
  & (Dare to) ask others questions    & \checkmark & \checkmark \\    
  & Dare to discover    &  & \checkmark \\    
\rot{\rlap{\textbf{How? (11)}}}  & Get out of comfort zone; be open to new things   & \checkmark & \checkmark  \\    
\midrule
  & Tooling support    & \checkmark & \checkmark \\    
  & Support from your environment    & \checkmark & \checkmark \\    
  & Being in a flow state    & \checkmark & \checkmark \\    
  & Individually behind the PC    & \checkmark &  \\    
  & In a (heterogeneous) group    & \checkmark & \checkmark \\    
  & In a quiet environment    & \checkmark & \checkmark \\    
  & Music/solitude    & \checkmark & \checkmark \\    
  & Sport/walking to take mind off of things    & \checkmark & \checkmark \\    
\rot{\rlap{\textbf{Where? (9)}}}  & Deliberate breaks to trigger "aha!" moment    & \checkmark &  \\    
\bottomrule
\end{tabular*}
\caption{Emerged subthemes and categories while analyzing transcripts of professionals (column P.) and students (column S.), after cross-validation steps. \label{fulltable}}
\end{center}
\end{table*}

\newpage

\bibliography{report}


\end{document}